# PREDICTION OF EXISTENCE OF NEUTRAL BOSON WITH SPIN 2 IN ENERGY (MASS) RANGE FROM ZERO TO 160.77 GEV

Vali A. Huseynov[1,2,3]

[1] Department of Theoretical Physics, Baku State University, Z. Khalilov 23, AZ 1148, Baku, Azerbaijan;
[2] Department of General and Theoretical Physics, Nakhchivan State University, University Campus, AZ 7012, Nakhchivan, Azerbaijan;
[3] Department of Physics, Qafqaz University, Baku-Sumgayit Road, 16 km., Khirdalan, Baku, AZ0101, Azerbaijan;  E-mail: vgusseinov@yahoo.com

**ABSTRACT**

We investigate the decay of an arbitrary neutral boson into a pair of on-shell $W$-bosons in a magnetic field. The possible existence of the new neutral bosons with the spins 0, 2, 3 and with the charge conjugation $C = +1$ in the energy (mass) range from zero to $160.77\,GeV$ is predicted. The analyses show that the existence of the neutral boson with the spin 2 in the energy (mass) range from zero to $160.77\,GeV$ is more promising and realistic.

Recently the ATLAS and CMS Collaborations reported on the discovery of a new neutral boson (NB) at a mass around $126\,GeV$ [1, 2] with properties compatible with the Standard Model Higgs boson [3-8] that is described with $J^{PC} = 0^{++}$ where $P$ is the parity, $C$ is the charge conjugation, $J$ is the spin. The production of new NBs in the $pp$-collisions in the energy (mass) range $0 < E < 2m_W$ ($m_W \cong 80.385\,GeV$ [9] is the $W$-boson mass) is not excluded in the future improved LHC or other experiments. Here we investigate the possibility of the existence of the new NBs with the spins 0, 2 and higher than 2 in the energy (mass) range $0 < E < 2m_W$ ($0 < E < 160.77\,GeV$). The following natural questions arise. Is the existence of any other NBs with the spin 0 except the $126\,GeV$ NB in the energy (mass) range $0 < E < 2m_W$ possible? How is realistic and promising the existence of the NB with the spin 2 and higher than 2 in the energy (mass) range $0 < E < 2m_W$? Search for the answers to these questions determines the motivation for the presented investigation. We investigate the decay of an arbitrary NB (we call it a neutral $Y$-boson) including the NB observed at the LHC into a $W^-W^+$-pair in a magnetic field (MF) provided that our arbitrary NB also decays into the two photons.

The main purpose of this work is to predict the existence of new NBs in the energy (mass) range $0 < E < 2m_W$ and to determine their spins.

Let us write the energy spectrum of a $W^{\mp}$-boson in a uniform MF [10, 11]

$$E_{W^{\mp}}^2 = p_{\mp z}^2 + (2n_{\mp} + 1 - 2q_{\mp}s_{\mp z})eB + m_W^2, \tag{1}$$

where $B = |\vec{B}|$ is the strength of a MF that is directed along the $Oz$-axis, $p_{\mp z}$ ($s_{\mp z}$) are the third component of the momentum (spin) of a $W^{\mp}$-boson, $n_{\mp} = 0, 1, 2, ...$ enumerates the Landau energy levels, $q_- = -1$ ($q_+ = +1$) is the sign of the electric charge of a $W^-(W^+)$-boson. A $W^{\mp}$-boson has three polarization states: $|W^{\mp}(s_{\mp} = 1, s_{\mp z} = +1)\rangle = |1, +1\rangle$, $|W^{\mp}(s_{\mp} = 1, s_{\mp z} = 0)\rangle = |1, 0\rangle$, $|W^{\mp}(s_{\mp} = 1, s_{\mp z} = -1)\rangle = |1, -1\rangle$, where $s_{\mp}$ is the spin of a $W^{\mp}$-boson. Hereafter we will consider the case $n_{\mp} = 0$, $p_{\mp z} = 0$ and $q_{\mp}s_{\mp z} = +1$ (i.e. the states $|W^-(s_- = 1, s_{-z} = -1)\rangle = |1, -1\rangle$ and $|W^+(s_+ = 1, s_{+z} = +1)\rangle = |1, +1\rangle$) that corresponds to the ground Landau level of the $W^-(W^+)$ -boson. When the $W^-(W^+)$-boson spin is oriented against (along) the MF direction (transverse polarization), i.e. when $s_{+z} = +1$ ($s_{-z} = -1$), the inequality $E_{W^{\mp}} = \sqrt{m_W^2 - eB} < m_W$ is satisfied for an arbitrary $B$ taken from the range $0 < B < B_{0W}$ where $B_{0W} = m_W^2/e$.

Let us consider the decay of an arbitrary NB into the $W^-W^+$-boson pair. For instance, the NB at a mass around $126\ GeV$ decays via the $H \to WW^* \to l\nu l\nu$ channel [12]. One of these $W^{\mp}$-bosons is on-shell, the other one ($W^*$) is off-shell. According to the energy conservation law the decay of this NB into the on-shell $W^-W^+$-boson pair is impossible. Therefore, this NB decays into one on-shell $W$-boson and one off-shell $W^*$-boson. However, if we place this NB or any other neutral $Y$-boson with the energy $E < 2m_W$ in a uniform MF, the MF will affect on the $W^-$- and $W^+$-bosons that are the products of the decay $Y \to W^-W^+$. If we have dealings with a massive NB, we can consider it in the rest frame. For the massive NB in the rest frame we have $E = E_{min} = m$. If we have dealings with a massless neutral boson (eg., graviton), we can not consider it in the rest frame and therefore its minimal energy $E_{min}$ will be determined by its wave frequency $\omega$, i.e. $E = E_{min} = E_{\omega}$. If we take into account the relation $E_{W^{\mp}} < m_W$ for the $W^-$ ($W^+$)-boson with $s_{-z} = -1$ ($s_{+z} = +1$) in the energy conservation law $E_{min} = E_{W^-} + E_{W^+}$, we can see that in a sufficiently strong MF with the strength $B_Y$ the equality $E_{min} = 2\sqrt{m_W^2 - eB_Y}$ satisfied and even the decay of any NB with the energy (mass) $E_{min} < 2m_W$ including the $126\,GeV$ NB into the two on-shell $W^{\mp}$-bosons becomes, in principle, energetically possible. So, as a result of the decay reaction $Y \to W^- W^+$ in a MF we have the final diboson system $Y'$ that consists of the on-shell $W^-$- and $W^+$-bosons situating in a MF. Here we disregarded the term $E_{amm}$ characterizing the interaction of the anomalous magnetic moment (if it exists) of the decaying NB with the MF on the left side of the equality $E_{min} = 2\sqrt{m_W^2 - eB_Y}$ and the term $E_C$ characterizing the Coulomb interaction energy between the on-shell $W^-$- and $W^+$-bosons on the right side.

We denote an arbitrary polarization state of the system $Y'$ as $|S_{Y'}, S_{Y'_z}\rangle$ where $S_{Y'} = S_{W^-W^+}$ is the total spin of the system $Y'$ and $S_{Y'_z}$ is the third component of the total spin vector $\vec{S_{Y'}} = \vec{S_{W^-W^+}}$. Using the two given polarization states $|W^-(s_- = 1, s_{-z} = -1)\rangle = |1, -1\rangle$ and $|W^+(s_+ = 1, s_{+z} = +1)\rangle = |1, +1\rangle$ of $W^{\mp}$-bosons, the addition rule of spins and the Clebsch-Gordan coefficients we obtain three different polarization states $|S_{Y'}, S_{Y'_z}\rangle$ of the system $Y'$: $|0,0\rangle, |1,0\rangle, |2,0\rangle$. So, the quantum states of the system $Y'$ can have the spin equal to 0, 1, 2. In the considered case the energy of the final $W^-W^+$-system $E_{W^-W^+}$ is in the range $0 < E_{W^-W^+} < 2m_W$ if we assume that the MF strength changes in the range $0 < B < B_{0W}$, where $E_{W^-W^+} = E_{W^-} + E_{W^+} = E_{min}$. Both for a massive NB and for a massless NB $E_{min} \neq 0$ and $E_{min} > 0$. Taking into account the condition $E_{min} \neq 0$ we obtain from the formula $E_{min} = 2\sqrt{m_W^2 - eB_Y}$ that $B_Y \neq B_{0W}$ and $B_Y < B_{0W}$. At the same time the presence of the external MF ($B_Y \neq 0$) means that in the considered case the minimal energy (mass) of the decaying neutral $Y$-boson also satisfies the condition $E_{min} < 2m_W$. Thus, if the NB at rest or massless NB decays in a MF into the two on-shell $W^{\mp}$-bosons producing on the ground Landau level with $n_{\mp} = 0$, $p_{\mp z} = 0$ and $q_{\mp} s_{\mp z} = +1$, it means that the minimal energy (mass) of this NB ought to be in the energy (mass) range $0 < E_{min} < 2m_W$. The NB at a mass around $126\,GeV$ is typical example for our arbitrary NB.

Introducing the intrinsic parity $P_{W^-}(P_{W^+})$ for the $W^-(W^+)$-boson, the orbital quantum number $L_{W^-W^+}$ and the total spin $S_{W^-W^+}$ for the $W^-W^+$-system we can determine the charge conjugation $C_{W^-W^+}$, the parity $P_{W^-W^+}$ and the total angular momentum $J$ of this system:

$$C_{W^-W^+} = (-1)^{L_{W^-W^+} + S_{W^-W^+}}, \qquad (2)$$

$$P_{W^-W^+} = (-1)^{L_{W^-W^+}} P_{W^-} P_{W^+} = (-1)^{L_{W^-W^+}}, \qquad (3)$$

$$J = L_{W^-W^+} + S_{W^-W^+}, L_{W^-W^+} + S_{W^-W^+} - 1, \ldots, |L_{W^-W^+} - S_{W^-W^+}|. \qquad (4)$$

We accept that the initial NB can also decay into the two photons. Therefore its spin $J$ can not be 1 according to the Landau-Yang theorem [13, 14] and the charge conjugation $C$ of the initial decaying neutral $Y$-boson is $C = C_Y = 1$. The decay $Y \to W^-W^+$ is a weak process and $C$ is not conserved in this process. It means that if the charge conjugation of the initial neutral $Y$-boson is $C_Y = 1$ before the reaction $Y \to W^-W^+$, the charge conjugation $C_{W^-W^+}$ might be $+1$ or $-1$ after the reaction, or it might also go to a state that is not a $C_{W^-W^+}$ eigenstate. Here we assume that $C_{W^-W^+}$ is either $+1$ or $-1$ after the reaction. We also assume that $P_{W^-W^+}$ is either $+1$ or $-1$ after the reaction. The following combinations of $C_{W^-W^+}$ and $P_{W^-W^+}$ for the $W^-W^+$-system are possible:

case A: $\qquad\qquad\qquad C_{W^-W^+} = +1, P_{W^-W^+} = +1,$ (5)

case B: $\qquad\qquad\qquad C_{W^-W^+} = +1, P_{W^-W^+} = -1,$ (6)

case C: $\qquad\qquad\qquad C_{W^-W^+} = -1, P_{W^-W^+} = +1,$ (7)

case D: $\qquad\qquad\qquad C_{W^-W^+} = -1, P_{W^-W^+} = -1.$ (8)

Taking into account the Landau-Yang theorem and using the relations (2)-(8) we have obtained that the NBs with the following spins can exist in the energy (mass) range $0 < E_{min} < 2m_W$) and they can decay into the on-shell $W^-(s_- = 1, s_{-z} = -1)$- and $W^+(s_+ = 1, s_{+z} = +1)$-bosons in a MF:

case A: if $C_{W^-W^+} = +1$ and $P_{W^-W^+} = +1$, $J = 0, 2$ ($S_{W^-W^+} = 0, 2$, $L_{W^-W^+} = 0$), (9)

case B: if $C_{W^-W^+} = +1$ and $P_{W^-W^+} = -1$, $J = 0, 2$ ($S_{W^-W^+} = 1$, $L_{W^-W^+} = 1$), (10)

case C: if $C_{W^-W^+} = -1$ and $P_{W^-W^+} = +1, J = 2, 3$ ($S_{W^-W^+} = 1$, $L_{W^-W^+} = 2$), (11)

case D: if $C_{W^-W^+} = -1$ and $P_{W^-W^+} = -1$, $J = 2$ ($S_{W^-W^+} = 2$, $L_{W^-W^+} = 1$) and

$$J = 3 \ (S_{W^-W^+} = 0, \ L_{W^-W^+} = 3; \ S_{W^-W^+} = 2, \ L_{W^-W^+} = 1). \quad (12)$$

We have obtained $J = 0, 2$ for the spin of the neutral $Y$-boson if $C_{W^-W^+} = +1$ and $J = 2, 3$ for the spin of the neutral $Y$-boson if $C_{W^-W^+} = -1$. One NB with the spin $J = 0$ has already been observed in the energy (mass) range $0 < E_{min} < 2m_W$ at the LHC [1, 2]. However the existence of the other NBs with the spin $J = 0$ and with the other masses is not excluded in the energy (mass) range $0 < E_{min} < 2m_W$. The analysis of the cases A, B, C and D show that the existence of the NB with the spin $J = 2$ is allowed in all possible cases A, B, C and D. When $W^-W^+$-pair are produced on the ground Landau level, the orbital quantum number $L_{W^-W^+}$ should be minimal. $L_{W^-W^+}$ is minimal only in the case A. So, the case A is more suitable for the particle with the spin $J = 2$. The energy (mass) of the NB with the spin $J = 2$ is in the energy (mass) range $0 < E_{min} < 2m_W$. If we use $J^{PC}$ assignment, for the NB with the spin $J = 2$ we have two possible cases: $2^{++}$ and $2^{-+}$. The existence of the NB with the spin 3 in the energy (mass) range $0 < E_{min} < 2m_W$ is not excluded, if $C_{W^-W^+} = -1$. The existence of the NB with the spin $J \geq 2$ would indicate that the world we live has additional dimensions besides known four ones [15, 16]. The MF strength required for the decay of an arbitrary neutral $Y$-boson with the energy (mass) in the range $0 < E_{min} < 2m_W$ into the on-shell $W^-(s_- = 1, s_{-z} = -1)$- and $W^+(s_+ = 1, s_{+z} = +1)$ is calculated by the formula $B_Y = B_{0W}[1 - (E_{min}/2m_W)^2]$. If we take into account the energy (mass) range $0 < E_{min} < 2m_W$ in the presented formula, the corresponding range for the MF strength will be given by the inequality $B_{0W} > B_Y > 0$. When $m \cong 126 \ GeV$ (the NB discovered at the LHC), the required MF strength is $\sim 10^{23} \ G$ (or in Teslas it is $\sim 10^{19} \ T$). If we assume $m \cong 159 GeV$, the required MF strength is $\sim 10^{22} \ G$. The maximum strength of the produced strong MF in noncentral heavy-ion collisions is estimated to be $\sim 10^{17} \ G$ at the RHIC and $\sim 10^{18} \ G$ at the LHC [17-23]. In lead-lead collisions at the LHC, the strength of the generated MF may reach $\sim 10^{20} \ G$ [18, 19]. We hope that in the future collider experiments, when the strength of the produced strong MF reaches the magnitude $\sim 10^{22 \div 23} \, G$, the decay of the NBs with the spins $J = 0, 2$ into the on shell $W^-$- and $W^+$-bosons can be observed experimentally. The predicted NBs with the spins $J = 0, 2$ can be also observed without a MF. The existence of the predicted NBs with the spins $J = 0, 2$ does not depend on the availability of the sufficiently strong MF. When a MF is absent, the NBs with the energy (mass) less than $2m_W$ and with the spins $J = 0, 2$ will decay into the one on-shell $W$-boson and one off-shell $W$-boson like $126 \, GeV$ boson.

So, we have obtained that the NBs with the spins $J = 0, 2, 3$ and with the charge conjugation $C = +1$ can exist in the energy (mass) range $0 < E_{min} < 2m_W$. The neutral $Y$-bosons with the spins $J = 0, 2$ ($J = 2, 3$) in the energy (mass) range $0 < E_{min} < 160.77 \, GeV$ is allowed if $C_{W^-W^+} = +1$ ($C_{W^-W^+} = -1$). The analyses of the obtained results enable us to come to the conclusion that the existence of the NB with the spin $J = 2$ and the charge conjugation $C = +1$ in the energy (mass) range $0 < E_{min} < 160.77 \, GeV$ is allowed in all possible cases. Therefore its existence is more promising and realistic. We hope that the possible existence of the NBs with the spins $J = 0, 2, 3$, especially, the existence of the new NB with the spin $J = 2$ in the energy (mass) range $0 < E_{min} < 160.77 \, GeV$ will attract the experimental physicists' attention in future collider experiments.

The author is very grateful to the Organizing Committee of the LHCP2014 Conference for the kind invitation to attend this conference.